\documentclass[preprint,prl,superscriptaddress,showpacs]{revtex4}
\usepackage{bm}
\usepackage{graphicx}

\def\URuSi{URu$_2$Si$_2$}
\def\jpcm{J.\ Phys.\ Condens.\ Matter\ }
\def\jmmm{J.\ Magn.\ Magn.\ Mater.\ }
\def\ssc{Solid State Commun.\ }
\def\jpsj{J.\ Phys.\ Soc.\ Jpn.\ }
\def\pb{Physica B\ }

\begin{document}

\title{Inflection point in the magnetic field dependence of the 
ordered moment of \URuSi\ observed by neutron scattering in fields up 
to 17 T}

\author{F. Bourdarot}
%\email{bourdarot@cea.fr}
\affiliation{D\'epartement de Recherche Fondamentale sur la Mati\`ere
Condens\'ee, SPSMS/MDN, CEA Grenoble, 38054 Grenoble, France}

\author{B. F\aa k}
\affiliation{D\'epartement de Recherche Fondamentale sur la Mati\`ere
Condens\'ee, SPSMS/MDN, CEA Grenoble, 38054 Grenoble, France}
\affiliation{ISIS Facility, Rutherford Appleton
Laboratory, Oxon OX11 0QX, England}

\author{K. Habicht}
\affiliation{Hahn Meitner Institut, BENSC, Glienicker Str. 100, 
D-14109 Berlin, Germany}

\author{K. Proke\v{s}}
\affiliation{Hahn Meitner Institut, BENSC, Glienicker Str. 100, 
D-14109 Berlin, Germany}

\date{12 August 2002; printed \today}

\begin{abstract}
We have measured the magnetic field dependence of the ordered 
antiferromagnetic moment and the magnetic excitations in the 
heavy-fermion superconductor \URuSi\ for fields up to 17 Tesla applied 
along the tetragonal $c$ axis, using neutron scattering.  The decrease 
of the magnetic intensity of the tiny moment with increasing field 
does not follow a simple power law, but shows a clear inflection 
point, indicating that the moment disappears first at the metamagnetic 
transition at $\sim$40 T. This suggests that the moment $m$ is 
connected to a hidden order parameter $\psi$ which belongs to the same 
irreducible representation breaking time-reversal symmetry.  The 
magnetic excitation gap at the antiferromagnetic zone center {\bf 
Q}=(1,0,0) increases continuously with increasing field, while that at 
{\bf Q}=(1.4,0,0) is nearly constant.  This field dependence is 
opposite to that of the gap extracted from specific-heat data.
\end{abstract}
\pacs{75.25.+z, 75.30.Kz, 75.30.Cr, 75.50.Ee, 74.70.Tx} 
\maketitle

One of the most enigmatic problems in heavy-fermion physics is to 
reconcile the smallness of the ordered antiferromagnetic (AFM) moment 
of \URuSi\ with the huge $\lambda$-anomaly in the specific heat, 
$C_p$, at $T_N=17.5$ K. Neutron 
\cite{broholm,mason1,walker,mason,fak,nse,honma} and resonant X-ray 
\cite{isaacs} scattering experiments have shown that a weak static 
antiferromagnetic dipolar order with the moment along the $c$ axis of 
the body-centered tetragonal structure (space group I4/mmm) of the 
heavy-fermion superconductor ($T_c=1.2$ K) \URuSi\ is established 
below $T_N$.  The magnitude of the moment (0.03 $\mu_B$/U-atom) is too 
small to reconcile with the jump in $C_p$.  The specific-heat anomaly 
suggests the opening of an energy gap over part of the Fermi surface 
\cite{palstra,dijk}.  Inelastic neutron-scattering 
measurements below $T_N$ show well-defined dispersive magnetic 
excitations polarized along the $c$ axis, which are consistent with a 
singlet-singlet crystal-field model \cite{broholm}.  However, 
such a model predicts an ordered moment much stronger than that 
observed.  Recent reports \cite{matsuda,amitsuka0} on that \URuSi\ 
samples are inhomogeneous with a larger moment in small regions are 
not yet experimentally confirmed and are still a matter of controversy 
and debate \cite{bernal}.  Also, they do not resolve the discrepancy 
between the small moment and the specific-heat anomaly.  

In this Letter we present the magnetic field dependence of the AFM 
component of the ordered moment and of the gaps in the magnetic 
excitation spectrum for fields $H$ up to 17 T applied along the $c$ 
axis.  The high magnetic fields now available for neutron scattering 
experiments at the HMI made it possible to observe that the magnetic 
moment in \URuSi\ does not vanish at $\sim$15 Tesla as was suggested 
from earlier measurements using a simple power-law extrapolation 
\cite{mason,santini}.  Instead, we observe a clear inflection point of 
the magnetic intensity near a field of 7 T and a finite moment at 17 
T. Such a behavior can be obtained if there is a linear coupling 
between a hidden primary order parameter and the magnetic moment 
\cite{shah}.  Inflection points have also been suggested in other 
theoretical schemes, which will be discussed.  The gap $\Delta$ in the 
magnetic excitation spectrum at the antiferromagnetic zone center {\bf 
Q}=(1,0,0) increases strongly with $H$.  The second gap $\nabla$ at 
{\bf Q}=(1.4,0,0) is nearly field independent.

Magnetic fields up to 17 T at low temperatures were obtained by 
mounting the sample between two dysprosium rods in a 14.5 T 
superconducting magnet \cite{Dy}.  A high-quality cylindrical \URuSi\ 
single crystal of diameter 6 mm and height 3.8 mm was cut by 
electroerosion from a larger well-characterized and annealed crystal 
already used in previous work \cite{santini} and mounted with the $c$ 
axis parallel to the vertical field of the magnet.  The neutron 
scattering experiment was made on the FLEX triple-axis spectrometer of 
HMI, using pyrolytic graphite as vertically focusing monochromator and 
horizontally focusing analyzer.  Measurement of the magnetic moment 
was done using the spectrometer in W (RLR) configuration with a wave 
vector of 1.48 {\AA}$^{-1}$.  To remove the higher-order harmonic 
contribution from the monochromator, we used a 10-cm thick 
liquid-nitrogen cooled Be filter in the incident beam and a graphite 
filter in the scattered beam, the latter oriented to reflect out the 
2\textit{nd} order harmonics.  The inelastic measurements were done 
with the spectrometer in a ``long-chair'' (RRL) configuration, a fixed 
final wave vector of 1.55 {\AA}$^{-1}$, and the Be filter in the 
scattered beam.  The energy resolution at elastic energy transfer was 
0.14 meV at full width at half maximum.  Most of the elastic 
and inelastic data were collected at a temperature of $T=2$ K for 
magnetic fields $H$ between 0 and 17 T.

\begin{figure}[p] % FIG. 1
\includegraphics[width=76mm]{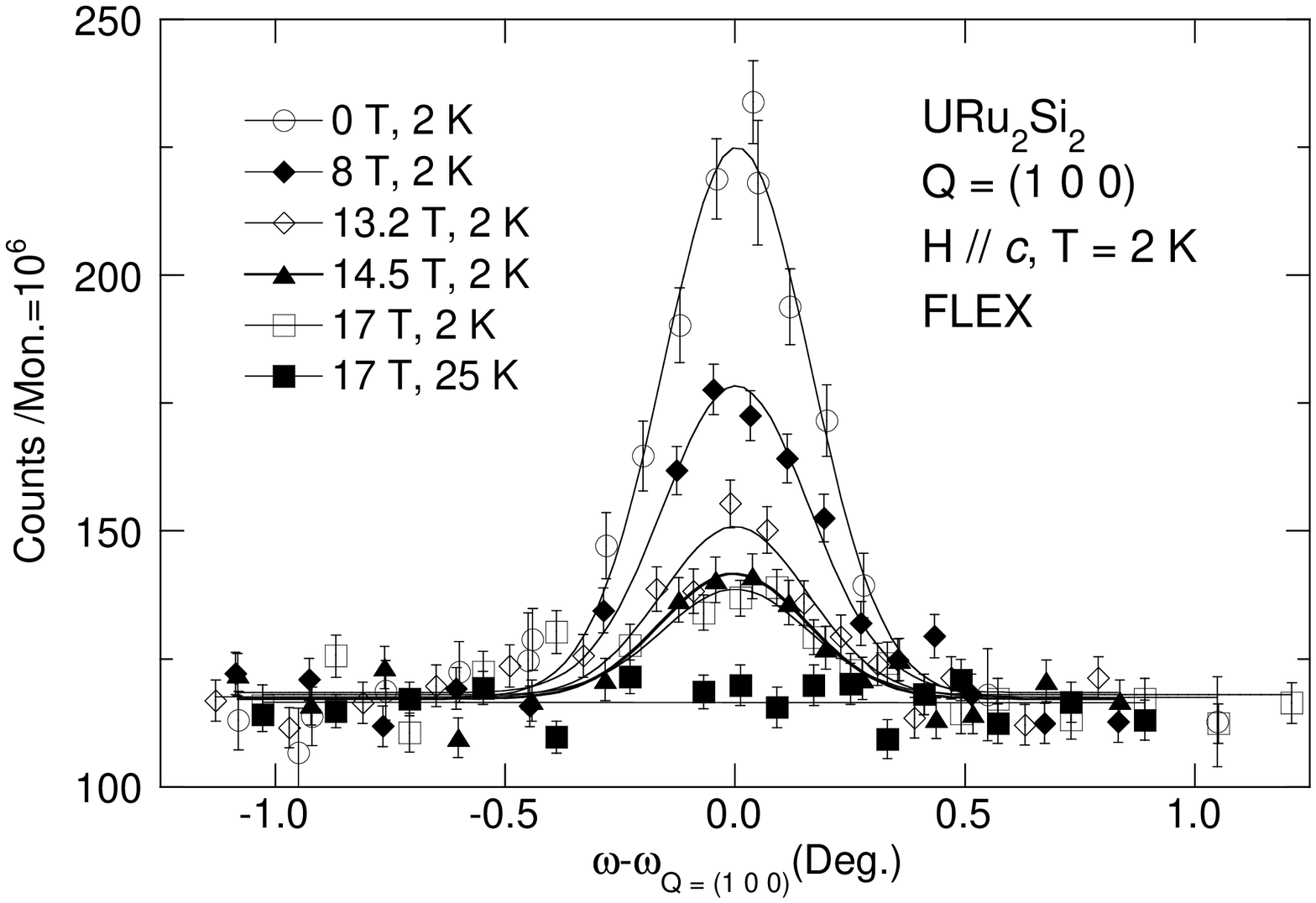}
\caption{Rocking scans of the AFM Bragg peak for different values of 
the magnetic field. The lines are Gaussian fits.}
\label{fig1}
\end{figure}

\begin{figure}[p] % FIG. 2
\includegraphics[width=76mm]{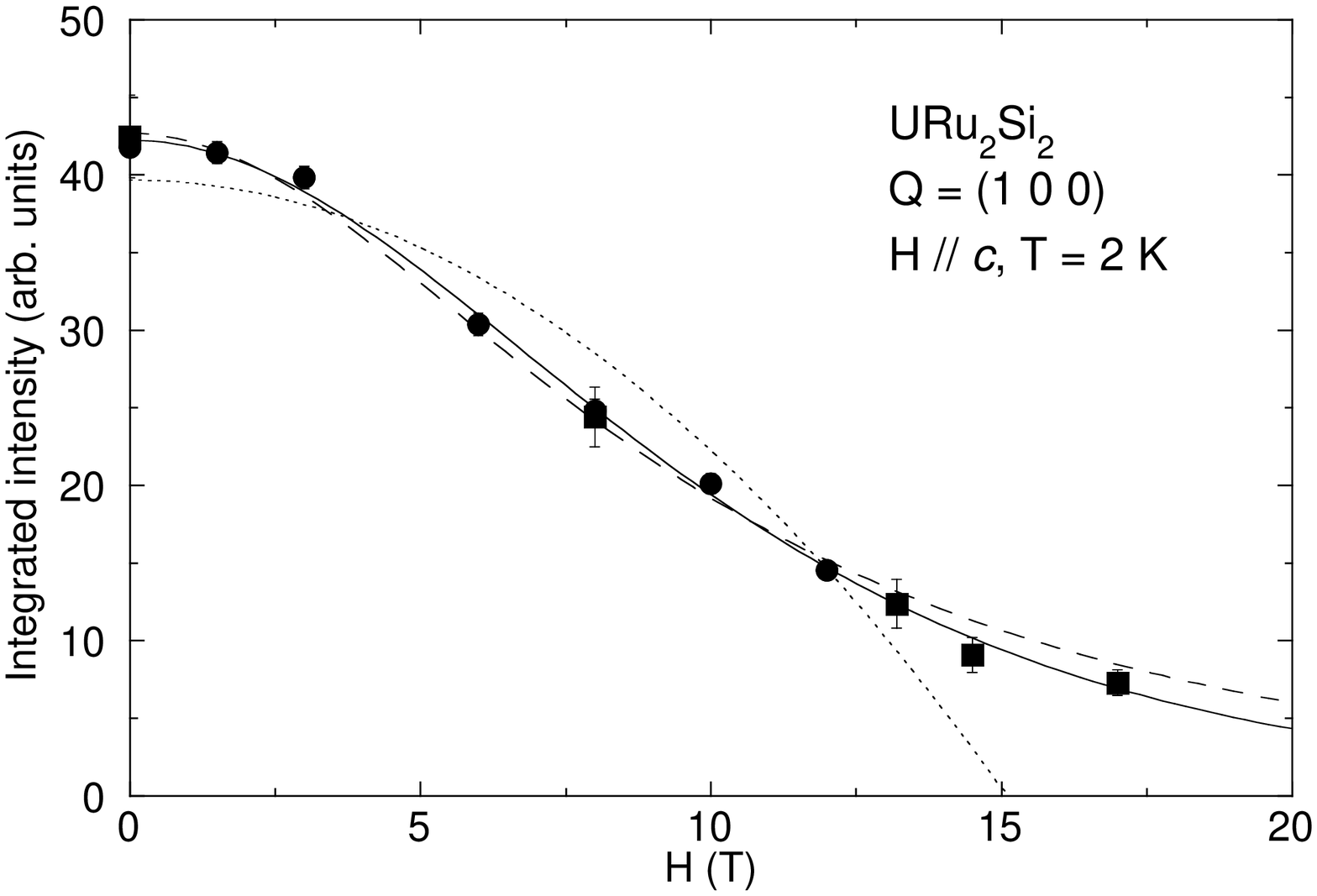}
\caption{Integrated intensity of the AFM Bragg peak as a function of 
magnetic field.  Squares (circles) are from FLEX (IN14).  The dotted, 
solid, and dashed lines correspond to fits given by Eqs.~(\ref{Eqm2}), 
(\ref{Shaheq}), and (\ref{Mineq}), respectively.}
\label{fig2}
\end{figure}

The integrated intensity of the AFM Bragg peak, which is proportional 
to the square of the ordered magnetic moment, $m^2$, was measured by 
rocking scans at {\bf Q}=(1,0,0), as illustrated in Fig.~\ref{fig1}.  
Measurements above $T_N$ (at high fields) show clearly the absence of 
any second-order contamination. The finite correlation length is 
field independent, as reported in earlier measurements \cite{santini}. 
Figure \ref{fig2} shows the 
integrated Bragg peak intensity as a function of applied field 
$H\parallel c$.  The figure combines the present data from HMI with 
earlier measurements at fields up to 12 T from the IN14 spectrometer 
at the ILL \cite{santini}, using the field points at 0 and 8 T to 
scale the results.  The magnetic moment decreases with increasing 
field, but clearly deviates at higher fields from the standard 
expression
\begin{equation}
   m^2 = m_0^2\: [1 - (H/H_m)^\gamma]
\label{Eqm2}
\end{equation}
with $\gamma = 2$.  Equation (\ref{Eqm2}) describes the magnetic field 
dependence at $T=0$ of a (single) order parameter $m$ (the magnetic 
moment) for a second-order phase transition, and is easily derived in 
Ginzburg-Landau theory using a free energy of the form
\begin{equation}
    F = \alpha(h^2-t)m^2+\beta m^4
\label{Free}
\end{equation}
with $t=1-T/T_N$ and $h=H/H_m$, where $T_N$ and $H_m$ are the critical 
temperature and field for the magnetic order, respectively, and 
$\alpha$ and $\beta$ are constants.  The critical field $H_m$ was 
extrapolated to $\sim$15 T in earlier low-field work 
\cite{mason,santini}.  Modifying the $\gamma$ exponent (a value of 3/2 
was used in Ref.~\cite{mason}) improves the fit but does not remove 
the disagreement with the data, which shows a clear inflection point at 
$\sim$7 T.

An inflection point in $m(H)$ was obtained in a model proposed by Shah 
{\it et al.} \cite{shah}, where the magnetic moment $m$ is a secondary 
order parameter, which is coupled linearly to a hidden primary order 
parameter $\psi$, which breaks time reversal symmetry.  The magnetic 
intensity can in this case be written as
\begin{equation}
m^2= m_0^2\:
\frac {t-(H/H_C)^2} {[1 + \delta(H/H_C)^2]^2},
\label{Shaheq}
\end{equation}
where $H_C$ is the metamagnetic field above which the heavy-fermion 
state is suppressed ($H_C$=35.8--39.4 T \cite{meta}; we use $H_C$=40 
T) and $\delta$ is a parameter.  Shah {\it et al.} introduces a 
characteristic field $H_M$ for the magnetic moment through 
$H_M=H_C/(1+2\delta)^{1/2}$.  Equation (\ref{Shaheq}) gives an 
excellent description of the observed field dependence, as illustrated 
in Fig.~\ref{fig2}, with $H_M = 10.4 \pm 0.2$ T as the only adjustable 
parameter.  The linear coupling term $\propto m\psi$ was originally 
suggested in Ref.~\onlinecite{agterberg}.  Symmetry arguments 
\cite{shah,agterberg} show that $\psi$ must have the same symmetry as 
$m$, i.e. $\psi$ belongs to the $A_{2g}$ irreducible representation (IR).  
The physical interpretation of $\psi$ is either in terms of 
triple-spin correlators (octupolar moments or couplings between 
dipolar and/or quadrupolar moments) or a dipolar moment of the 
conduction electrons, which have such a rapidly decreasing form factor 
that the magnetic intensity is negligibly small already at the first 
observable magnetic Bragg peak \cite{agterberg}. 

The present finding of an inflection point in $m(H)$ rules out the 
possibility that the primary order parameter is even under time 
reversal.  Taking the simplest coupling term $m^2\psi^2$ with the 
right AFM symmetry \cite{shah,agterberg}, the field dependence of $m$ 
is the same as in Eq.~(\ref{Eqm2}) \cite{shah}, and hence in 
contradiction with the present results.  Recent $^{21}$Si NMR data suggest 
in fact
that $\psi$ breaks time-reversal symmetry \cite{bernal}. 
The existence of an inflection point in $m(H)$ also excludes the scenario 
with two decoupled order parameters and accidentally close transition 
temperatures \cite{paolo}, since $m(H)$ would then follow Eq.~(\ref{Eqm2}). 

While a hidden order parameter that breaks time-reversal symmetry is 
the most likely scenario, an inflection point in $m(H)$ can also be 
obtained in a strong coupling model with the magnetic moment as the 
only order parameter \cite{mineev}.  If the total moment changes 
between the normal and the magnetically ordered state due to an 
increase of the magnetic coupling below $T_N$, then the $\beta$ 
term will depend on the external 
field $h$ and is replaced by $\beta(1+\eta h^2)$ in Eq.~(\ref{Free}), 
which leads to
\begin{equation}
    m^2 = m_0^2\: \frac{1 - h^2}{1+\eta h^2}.
\label{Mineq}
\end{equation}
Equation~(\ref{Mineq}) provides a good description of the present 
neutron data, with $\eta = 17.4 \pm 0.7$, as seen in Fig.  \ref{fig2}.  
However, the model does not resolve the discrepancy between the small 
moment and the specific-heat anomaly.

\begin{figure}[p] % FIG. 3 a+b
\includegraphics[width=76mm]{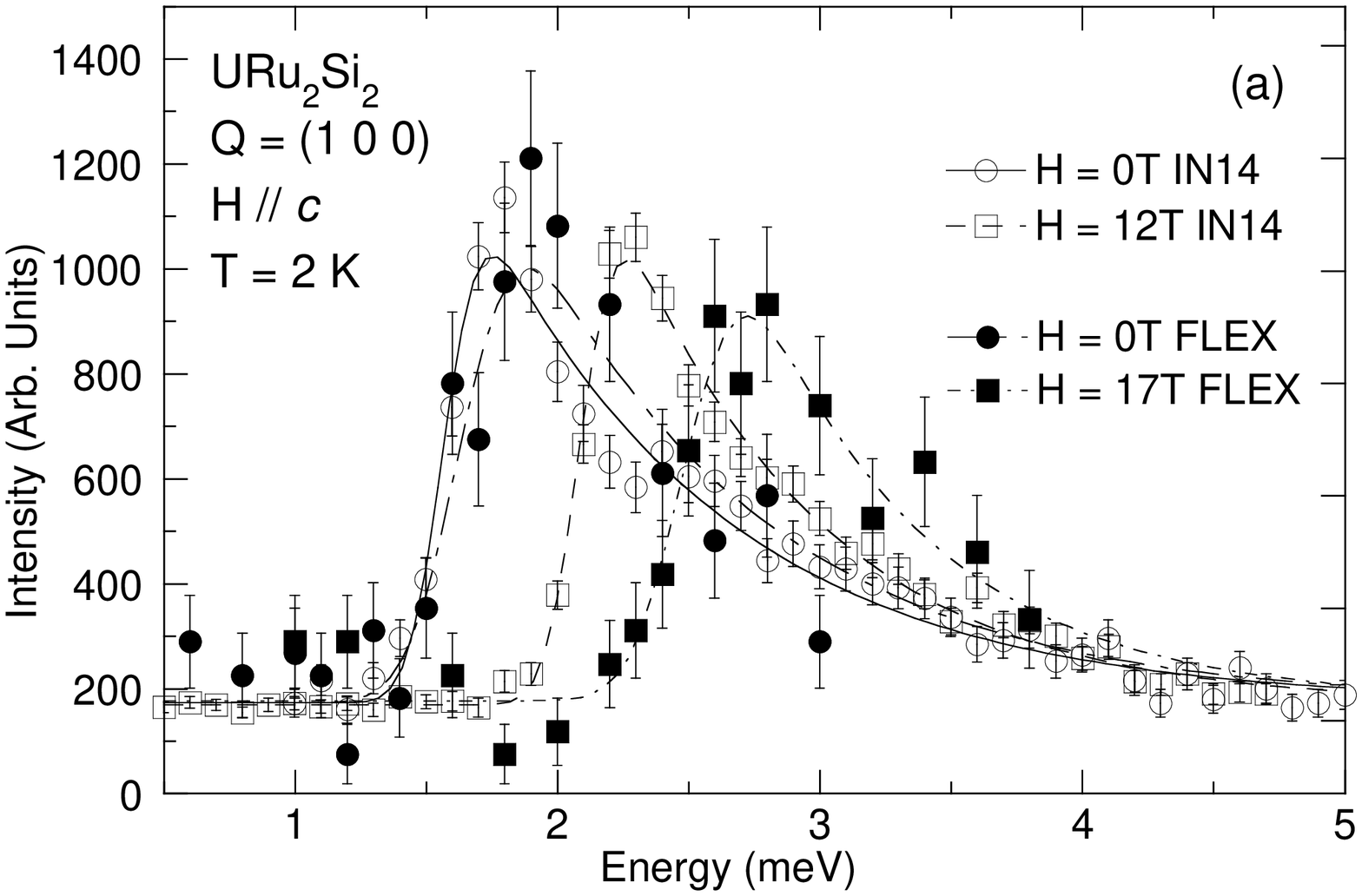}
\includegraphics[width=76mm]{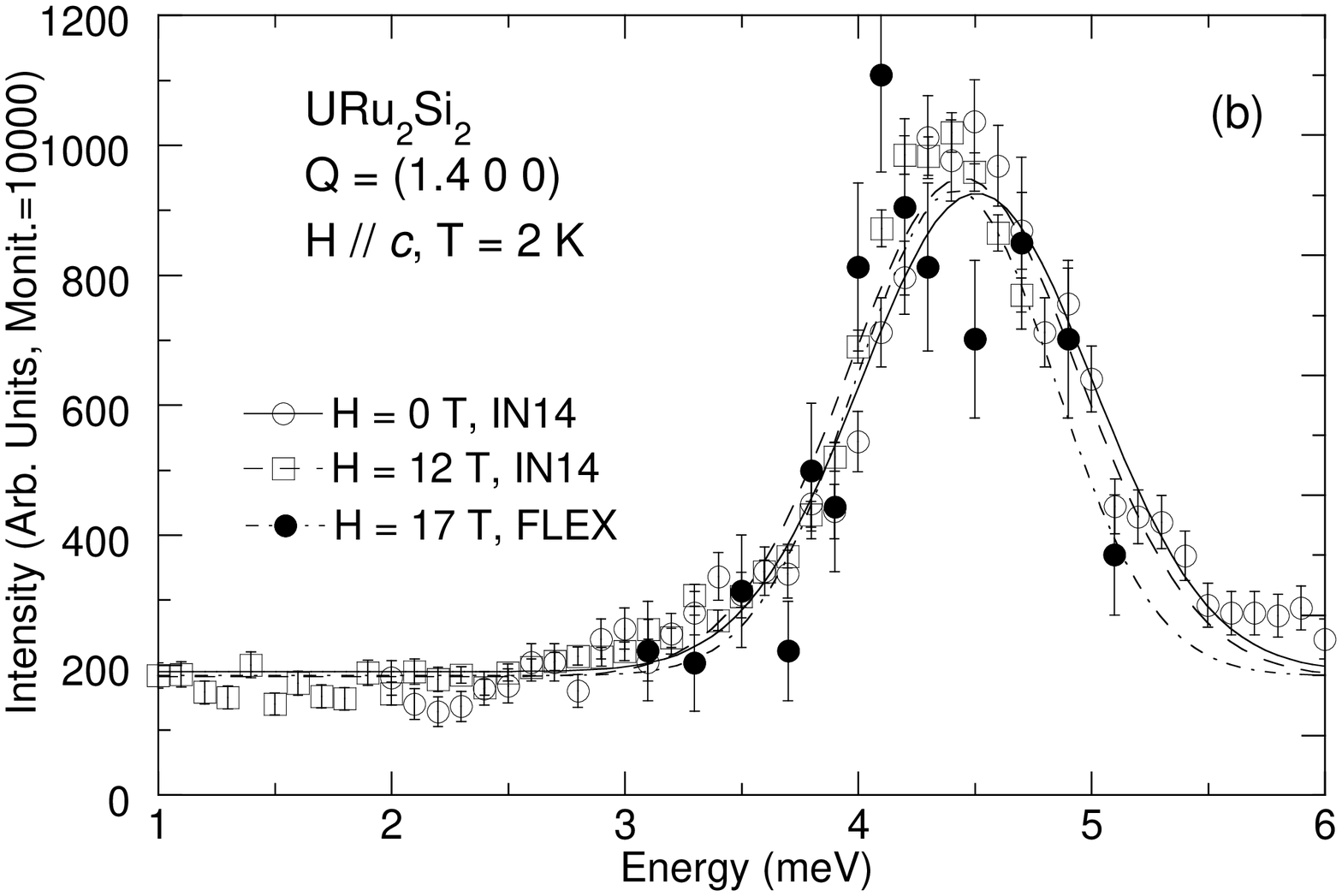}
\caption{Energy scans of the magnetic excitations at the two local 
minima {\bf Q}=(1,0,0) (a) and {\bf Q}=(1.4,0,0) (b) for magnetic 
fields of 0, 12, and 17 T.  The intensity scales for the data on the 
IN14 and FLEX spectrometers are scaled to coincide.  The lines are 
fits described in the text.  }
\label{new3}
\end{figure}

Okuno and Miyake \cite{okuno} have proposed a model which combines the 
localized character of a the singlet-singlet crystal-field scheme with 
an itinerant character of the 5$f$ electrons.  They assume further 
that there is a perfect nesting at the AFM zone center {\bf 
Q}=(1,0,0).  For a suitable parameter set, they obtain an inflection 
point in the field dependence of the magnetic moment whose shape 
resembles that of the present results.  However, for the parameter 
sets they employ, the inflection point occurs at magnetic fields 
several times higher than the observed inflection point and it 
seems difficult to obtain a semi-quantitative 
agreement with experimental values of the size of the ordered moment, 
the transition temperature, and the inflection point, simultaneously.

We note in passing that a magnetic field applied along the $c$ axis 
induces a localized 
ferromagnetic component, resulting in a ferrimagnetic structure with 
two different moments on the uranium-ion sites.  The magnitude of the 
ferromagnetic moment at the metamagnetic transition can be estimated 
to 0.32 $\mu_{B}$, based on an extrapolation of low-field polarized 
neutron diffraction measurements \cite{schweizer}.

The magnetic excitations at the local minima {\bf Q}=(1,0,0) and 
(1.4,0,0) of the dispersion curve have been measured at low 
temperature ($T=2$ K) for magnetic fields up to 17 T using the FLEX 
spectrometer of HMI. These data have been combined with earlier 
measurements at $H\leq 12$ T using the IN14 spectrometer of ILL 
\cite{santini}.  The results are shown in Fig.~\ref{new3}.  The 
measurements on IN14 have better statistics, due to higher flux and a 
larger sample volume (the sample volume is very limited in the 17 T 
magnet).  Instrumental resolution effects in combination with the 
steep dispersion introduce an asymmetric high-energy tail of the sharp 
excitation at the magnetic zone center {\bf Q}=(1,0,0).  In such a 
case, the excitation energy can be extracted by fitting the 
convolution of a Gaussian function, which describes the energy 
resolution, with an exponential saw tooth function 
$\exp[-(E-\Delta)/\epsilon]$, which describes how the higher-energy 
excitations are coupled into the measured spectrum via the finite $Q$ 
resolution.  The field dependence of the parameter $\epsilon$, which 
has no physical meaning, was determined from measurements at high 
energy transfers on IN14, and extrapolated to higher field values.  
This method gives a very accurate description of the measured spectra, 
as shown in Fig.~\ref{new3}(a), and the obtained excitation energy 
$\Delta$ lies, as expected, at approximately the half-height of the 
leading (low-energy) edge of the measured intensity.  The peak shape 
of the excitation at the second minima, {\bf Q}=(1.4,0,0), is 
symmetric [see Fig.~\ref{new3}(b)] because of the smaller dispersion, 
and a Gaussian fit is sufficient to extract the corresponding 
excitation energy $\nabla$.

\begin{figure}[p] % FIG. 4
\includegraphics[width=70mm]{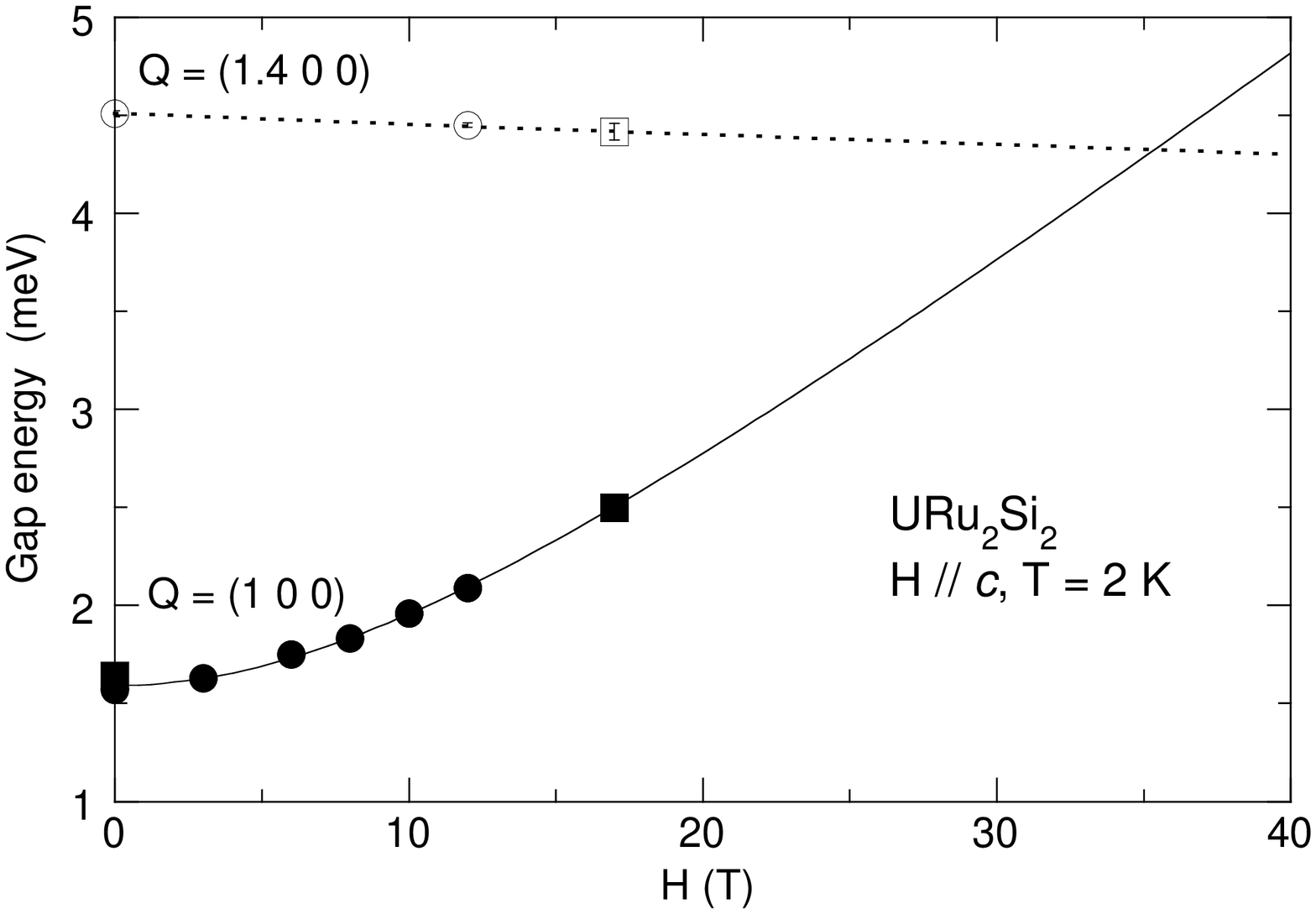}
\caption{The gap energies at {\bf Q}=(1,0,0) (solid symbols) 
and {\bf Q}=(1.4,0,0) (open symbols) as a function of magnetic field 
measured on IN14 (circles) and FLEX (squares).  The solid line is a fit 
to Eq.~(\ref{EqGap}) and the dotted line is a linear fit.  }
\label{new4}
\end{figure}

The magnetic field dependence of the gap energies $\Delta$ and 
$\nabla$ are shown in Fig.~\ref{new4}.  The gap at the AFM zone center 
increases rather strongly with applied field, without any sign of 
saturation at higher fields.  It is well described by
\begin{equation}
\Delta(H)=[\Delta_0^2+(bH)^2]^{1/2}
\label{EqGap}
\end{equation}
with $\Delta_0=1.59$ (2) meV and $b=0.114$ (1) meV/T. The gap 
$\nabla=4.51$ (2) meV at {\bf 
Q}=(1.4,0,0) is nearly constant, with only a slight linear decrease in 
energy with increasing field, with slope 5.4 $\pm$ 2.7 $\mu$eV/T. It 
is interesting to note that the high-field extrapolation of the energy 
gaps $\Delta(H)$ and $\nabla(H)$ crosses at a field close to the 
metamagnetic transition $H_C$.  This is very interesting in the light 
of the appearance of a new high-field phase above 36.1 T \cite{jaime}, 
which could be due to Fermi surface effects requiring a different 
nesting vector, such as the one corresponding to {\bf Q}=(1.4,0,0).  

Specific heat \cite{dijk} and electrical resistivity \cite{mentink} 
data taken with a magnetic field $H\parallel c$ up to 17.5 and 16 T, 
respectively, have been analyzed in terms of a field-dependent energy 
gap related to the excitation gap observed in neutron scattering 
measurements.  In both cases, the N\'eel temperature and the gap 
energy decrease quadratically with the field, following the 
relations $T_N(H)=T_N(0)[1-(H/H_0)^2]$ and 
$\Delta(H)=\Delta(0)[1-(H/H_0)^2]$, respectively, with $H_0$ being 
close to the metamagnetic field.  Very recent specific-heat 
measurements at fields up to 45 T \cite{jaime} show that  $T_N$ vanishes 
at a critical field of 35.9 T. The field 
dependence of $T_N$ is essentially quadratic.  Thermal expansion data 
\cite{vistink} with a magnetic field $H\parallel c$ up to 25 T have 
been fitted by the empirical function $A\exp(-\Delta/T)$ for the 
magnetic contribution.  A quadratic decrease of $\Delta$ and $T_N$ was 
found with an extrapolated critical field of 37 T and a gap energy 
of 115 K at zero field.

The important point is that the energy gap extracted from different 
bulk measurements has the opposite magnetic field dependence to the 
gap $\Delta$ observed directly by neutron scattering at the 
AFM zone center.  This low-energy gap actually {\em 
increases} strongly with field.  The second gap $\nabla$ shows a very 
slight linear decrease, much slower than a quadratic decrease observed 
in macroscopic measurements.  While specific heat and electrical 
resistivity measure some average gap energy, it is difficult to 
imagine how an average of a strongly increasing low-energy gap and a 
nearly field-independent high-energy gap would give rise to a 
decreasing gap.  It is therefore clear that the gap from bulk 
measurements is unrelated to the gap(s) in the magnetic excitation 
spectra.  This is in stark contrast to the transition temperature, 
which has the same field dependence in specific heat 
\cite{vistink,dijk,jaime}, electrical resistivity 
\cite{mentink,jaime}, thermal expansion \cite{vistink}, and neutron 
scattering \cite{mason,in8} measurements.  Unfortunately, due to the 
difficulty to measure $T_N$ accurately for small moments with neutron 
scattering, there are no systematic neutron measurements of $T_N(H)$. 
It is interesting to note that the height of the anomalies in the specific 
heat $\Delta C_p$ \cite{dijk} and resistivity $\Delta\rho$ \cite{mentink} 
as well as the intensity of the magnetic gaps $\Delta$ 
and $\nabla$ (see Fig. \ref{new3}) are field independent.

In conclusion, our neutron scattering measurements on the 
heavy-fermion superconductor \URuSi\ show that the magnetic field 
dependence of the antiferromagnetically ordered moment $m$ has a clear 
inflection point at $\sim 7$ T and remains finite even at 17 T for 
fields applied along the $c$ axis.  This strongly suggests the 
existence of a hidden order parameter $\psi$ that breaks time-reversal 
symmetry.  Symmetry arguments show that the hidden order parameter 
belongs to the same irreducible representation $A_{2g}$ as $m$ and 
corresponds to (ordinary) dipolar moments of the {\it conduction 
electrons} or triple-spin correlators of the 5$f$ electrons.  Our 
neutron scattering measurements of the (gapped) magnetic excitations 
show that they are {\it not} related to the gap extracted from e.g. 
specific heat measurements, since the magnetic field dependence of 
these gaps is completely different.  Instead, it is the hidden order 
parameter that gives rise to the large anomaly in specific heat and 
other macroscopic properties.  The extrapolated crossing of the two 
magnetic gaps might explain the new phase recently discovered 
\cite{jaime} above the metamagnetic transition.

We acknowledge fruitful discussions with 
G. Amoretti, N.H. van Dijk, J. Flouquet, I. Fomin, V.P. Mineev, K. 
Miyake, P. Santini, J. Schweizer, M.B. Walker, and M.E. Zhitomirsky.
A.A.  Menovsky, P. Lejay, and A.D. Huxley contributed to sample growth 
and preparation.
We thank M. Jaime for giving us access to his high-field 
specific-heat data prior to publication.

\end{document}